\documentclass[10pt,letterpaper]{article}
\usepackage{opex3}
\usepackage{graphicx}
\usepackage{amsmath,amssymb}

\begin{document}
\title{Narrow-band single-photon emission in the near infrared for quantum key distribution}
\author{E Wu and Vincent Jacques}
\address{Laboratoire de Photonique Quantique et Mol\'eculaire, UMR CNRS 8537, Ecole Normale Sup\'erieure de Cachan, France}
\author{Heping Zeng}
\address{Key Laboratory of Optical and Magnetic Resonance Spectroscopy, East China Normal Univerity, Shanghai 200062, P.R. China}
\author{Philippe Grangier}
\address{Laboratoire Charles Fabry de l'Institut d'Optique, UMR CNRS 8501, France} 
\author{Fran\c cois Treussart and Jean-Fran\c cois Roch}
\address{Laboratoire de Photonique Quantique et Mol\'eculaire, UMR CNRS 8537, Ecole Normale Sup\'erieure de Cachan, France}

\email{francois.treussart@physique.ens-cachan.fr}

\begin{abstract}
We report on the observation of single colour centers in natural diamond samples emitting in the near infrared region when optically excited. Photoluminescence of these single emitters have several striking features, such as a narrow-band fully polarized emission (FWHM 2~nm) around $780$~nm, a short excited-state lifetime of about $2$~ns, and perfect photostability at room temperature under our excitation conditions. We present a detailed study of their photophysical properties. Development of a triggered single-photon source relying on this single colour centre is discussed in the prospect of its application to quantum key distribution.
\end{abstract}
 

\section{Introduction} 

Significant efforts have been expended recently in the development of new and reliable single-photon sources, due to an increasing number of applications of such sources in quantum optics, namely  in quantum cryptography and for the realization of quantum gates~\cite{NJP_special_issue_SPS}. For instance, basic security proofs of quantum key distribution (QKD) rely on encoding information on single quantum objects, namely single-photon light pulses~\cite{BB84, Lutkenhaus_PRA99}.
Most practical QKD systems use weak coherent pulses as an approximation of single-photon sources, which greatly simplifies the experimental implementation~\cite{Gisin}. However, faint laser pulses with mean photon number smaller than unity can still contain more than one photon due to Poissonian photon number statistics. Increasing attenuation on the quantum channel to lower the probability of multiphoton pulses is a practical solution, but it also reduces the secret key creation rate. Moreover, due to dark count noise of the detectors, the vulnerability of the most commonly used QKD Bennett and Brassard BB84 protocol~\cite{BB84} increases with losses on the transmission channel~\cite{Lutkenhaus_PRA99}. To improve security of long-distance QKD, one as either to conceive of new protocols, like the one relying on decoy states~\cite{Hwang_PRL_03}, or to switch to single-photon sources.

Single photons can be conditionally prepared using a pair emission process like atomic cascade~\cite{Grangier_EPL86} or spontaneous parametric down-conversion~\cite{Hong}. One photon of the emitted pair is used as a ``trigger'' which prepares the field modes corresponding to the second photon in a single-photon state.
With this scheme, heralded single photons at telecom wavelengths can be efficiently prepared using bulk $\chi^{(2)}$ nonlinear crystals~\cite{Fasel_NJP04}, or guided-wave nonlinear devices~\cite{Alibart_OptLett05}. These systems have been experimentally confirmed as promising for optical fiber-based long-distance QKD~\cite{RomainAIP} .

However, emission time remains random for this type of single-photon generation. In quantum cryptography-based communication system, it is highly desirable to deal with single photons synchronized on an external clock.

Clock-triggered single-photons can be efficiently produced by suited pulsed excitation of the emitting dipole~\cite{Brouri}, which should undergo a full cycle, including excitation, emission and reexcitation before it emits the next photon. Since first proposal by De Martini {\it et al.}~\cite{DeMartini96} a wide variety of schemes to implement a single-photon source have been worked out, relying on the fluorescence of an isolated single emitter, such as a molecule, an atom, a colour centre, and a semiconductor quantum dot~(see Ref.\cite{NJP_special_issue_SPS} for a review).

A reliable room-temperature clock-triggered single-photon source was recently realized, based on the pulsed optically excited photoluminescence of a single nitrogen-vacancy (NV) colour centre in a diamond nanocrystal~\cite{Alexios_EPJD}. This turn-key single-photon source was then implemented in QKD experiments using BB84 protocol with polarization-encoded single-photon sequences~\cite{Alexios_PRL02,Romain}. As theoretically expected~\cite{Lutkenhaus_PRA99}, use of single-photon light pulses showed measurable advantages over faint laser pulses in terms of security of the quantum channel and key creation rate in the high attenuation regime.

The wide variety of photostable colour centres in diamond~\cite{Zaitsev} offers high opportunity to optimize single-photon properties so that they better meet application requirements for practical QKD. For instance, polarized single-photon emission, together with nanosecond-range excited state lifetime and nanometer spectral bandwidth, should be achieved. Recently, the nickel-related NE8 defect was observed at the single emitter level, both in natural diamond sample~\cite{Gaebel} and in CVD diamond film~\cite{Rabeau}. This colour centre, which consists in a nickel atom in divacancy position and four nitrogen atoms in its first coordination sphere, is characterized by a weak electron-phonon coupling. Most of its photoluminescence is then concentrated within its zero-phonon line, corresponding to a sharp emission line (FWHM 1.5 nm) around 800~nm.

We report on  the observation at room temperature of single diamond colour centres with perfect photostable luminescence  at room temperature, in the same wavelength range as the NE8 colour centre. Assumed to be also nickel-nitrogen related impurities, we study in  details their photoluminescence properties under optical excitation with a red cw laser diode. Both single-photon emission and shelving in a metastable level are observed from photon time-correlation measurements, with an Hanbury Brown \& Twiss setup. We show that these emitters meet the required properties for practical single-photon QKD as stated above. First, a narrow emission band is observed around 782~nm, which is both in the open-air and fiber-optics optical communication window. Second, an excited-level lifetime around 2~ns is inferred, which is compatible with narrow time-window analysis in order to reduce the effect of photodetector dark count noise. Third, light emitted is fully linearly polarized, which is perfectly suitable for polarization information encoding on the emitted single photons.
 
\section{Experimental set-up and results}
\begin{figure}[htbp]
\begin{center}
\includegraphics[width=0.9\textwidth]{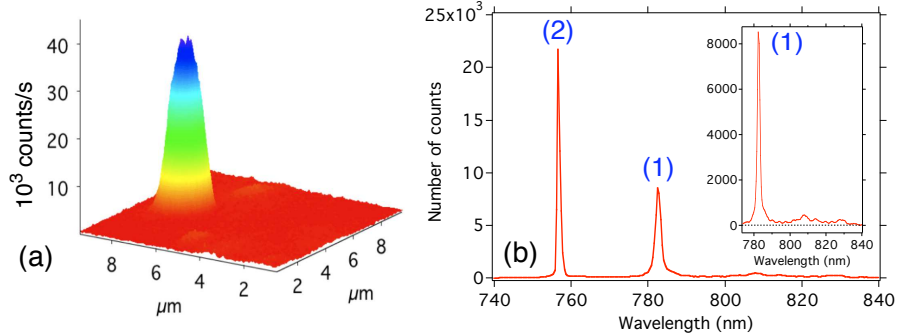}
\caption{(a) Fluorescence intensity raster scan of a natural diamond sample showing luminescence from an isolated colour centre. The counting rate corresponds to the output of one detector in the HBT setup. Its maximum value is about $4\times 10^4$~counts/s. (b) Photoluminescence spectrum from the single emitter observed in Fig.\ref{scan}(a) with 10~s integration duration. The spectrum displayed has been corrected for the quantum efficiency of the CCD which varies from 85 to 53\% in the 740--840~nm range considered. The narrow peak (1) at 782~nm (about 2~nm FWHM) is the zero-phonon line of the nickel-nitrogen related defect~\cite{Gaebel}. The sharp peak (2) at 756~nm is related to the one-phonon Raman scattering line of the diamond lattice  with 1332~cm$^{-1}$ frequency shift. \emph{Inset}: zoom on the zero-phonon line, showing more clearly the intensity of the phonon wing.}
\label{scan}
\end{center}
\end{figure}

We use an home-made scanning confocal microscope to select the single defects in the sample (see Ref.\cite{Rosa00} for a detailed description of the setup). The detection system includes a Hanbury Brown \& Twiss (HBT) setup with two silicon avalanche photodiodes (APDs) in single-photon counting regime (SPCM, Perkin Elmer) on each side of a 50/50 non-polarizing beam splitter. It also integrates a spectrograph with a cooled CCD array for recording spectrum.
We first raster scan the sample ($2\times2\times0.32$~mm type IIa natural bulk diamond wafer from Element6, The Netherlands) to identify well isolated photoluminescent emitters using a laser diode emitting at $687$ nm as  cw excitation source. The laser beam is focused on the sample about $4$ $\mu$m below its surface by a microscope objective ($\times100$, NA$=0.95$) which is also used for emitted light collection. The laser focus spot is about $1\mu$m FWHM. Fig.\ref{scan}(a) displays a $9\times 9$ $\mu$m scan of the sample,  showing photoluminescence from a single nickel-nitrogen related defect. The $z$-axis value indicates the count rate of one APD of the HBT setup. From the scan in Fig.\ref{scan}(a), the observed signal to background ratio of the photoluminescence is as large as $60:1$. 

The HBT setup is used to investigate the fluorescence intensity correlation function. Thanks to the narrow spectral emission, a bandpass filter placed before the HBT setup allows us to select the photoluminescence from the defect while rejecting Raman scattered light due to the diamond crystalline lattice. Appropriate data processing of the photodetection events in the HBT configuration gives the histogram of the time separations between two consecutively detected photons. This delay function is equivalent to the second-order intensity autocorrelation function for short time scales~\cite{Reynaud_these_etat}. Evidence for single centre emission is given by  the observation of antibunching~\cite{Knight_book_05}. This effect appears as a dip around $t = 0$ in the recorded delay function (see Fig.\ref{fig_g2_short}(a)), as the consequence of finite-time cycle of excitation-emission-reexcitation for a single dipole. 

For each observed single emitter, we simultaneously record the spectrum of the emitted light on the spectrograph. Fig.\ref{scan}(b) shows the spectrum of the light collected from Fig.1(a) emitter. Narrow emission peak at 782~nm is tentatively associated to a nickel-nitrogen related impurity in diamond~\cite{Zaitsev,Yelisseyev}. From the spectrum, we infer that even at room temperature about 70\% of the total intensity is concentrated in the zero-phonon line. This value is much higher than that of nitrogen-vacancy colour centre, for which it is only of the order of 1\%.

We also investigated the polarization properties of the single colour centre relative to absorption and emission of light. We have monitored the photoluminescence intensity while rotating the excitation laser linear polarization angle and obtained oscillations with a contrast of about 96\%. This contrast value proves that the single colour centre behaves like a dipole relative to absorption of light. Using the quarter wave plate method, we studied the emitted light polarization properties. We observed that the collected light is perfectly elliptically polarized with an aspect ratio of 0.25. This measurement indicates that the single defect also behaves like an emitting dipole. The linear polarized light becomes elliptical after propagation though optics including high numerical aperture objective and dichroic mirror, both known for modifying the polarization state of light. 

\section{Photophysics of the 782~nm emitting colour centres}
\begin{figure}[htbp]
\begin{center}
\includegraphics[width=0.6\textwidth]{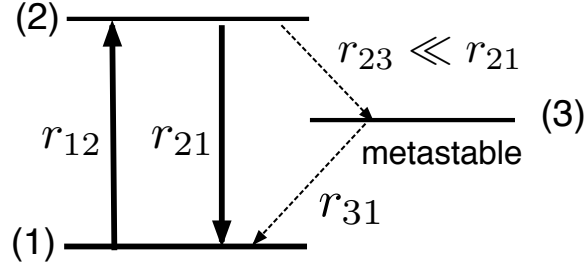}
\caption{Three energy levels scheme for the single colour centre emitting at 782~nm central wavelength, with corresponding decay and intersystem crossing rates.}
\label{level_diag}
\end{center}
\end{figure}

In order to investigate photoluminescence dynamical processes, we model the response of the single colour center within the framework of a three-level system corresponding respectively to ground level (1), excited level (2), and metastable level (3) as shown in Fig.\ref{level_diag}. 

The single colour centre is excited from its ground to excited level. It then either goes back to the ground  level, or it non-radiatively decays to the metastable level, which presence is asserted by photon bunching in the intensity correlation measurement. From this ``dark'' level, the system finally goes back to the ground level by a phosphorescence process. The dynamics can be described by a set of rate equations on the populations $\{p_1,p_2,p_3\}$ of the three considered levels
\begin{equation}
\left\{
\begin{array}{ll}
\dot{p_1}=&-r_{12}p_1+r_{21}p_2+r_{31}p_3\\
\dot{p_2}=&r_{12}p_1-(r_{21}+r_{23})p_2\\
\dot{p_3}=&r_{23}p_2-r_{31}p_3
\end{array}
\right.
\end{equation}
where $r_{nm}$ ($n, m=1,2,3$) denotes the transition rate from the level $(n)$ to level $(m)$. We assume that the transition rates $r_{23}$ and $r_{31}$ respectively to and from the metastable level are small compared to the excited leve to the ground-level decay rate $r_{21}$. 
Following Ref.~\cite {Kitson}, we can infer, for the normalized intensity correlation function
\begin{equation}
g^{(2)}(t)\equiv \frac{\langle I(0)I(t)\rangle}{\langle I(t)\rangle^2}
\end{equation}
the following analytical expression, valid for $t\ge 0$
\begin{align}
g^{(2)}(t)={}&p_2(0;t)/p_2(\infty)=1-(1+a)e^{-\lambda_{1}t}+ae^{-\lambda_{2}t},\label{g2}\\
\lambda_{1}={}&r_{12}+r_{21},\label{lambda1}\\
\lambda_{2}={}&r_{31}+r_{23}r_{12}/(r_{12}+r_{21}),\label{lambda2}\\
a={}&r_{12}r_{23}/[r_{31}(r_{12}+r_{21})],
\end{align}
where $p_2(\infty)$ is the steady-state population of the excited level (2), and where $p_2(0;t)$ denotes the population of level (2) at time $t$, starting from level (1) at time $t=0$. Recalling that $r_{23}$ and $r_{31}$ are small compared to $r_{21}$, it follows that $\lambda_1$ is much larger than $\lambda_2$. Two limit expressions of Eq.\ref{g2} can then be found

\noindent (i) On ``short'' time scale ($t\lesssim 20$~ns), the $g^{(2)}$ function exhibits photon antibunching:
\begin{equation}
g^{(2)}(t)\simeq1-(1+a)e^{-\lambda_{1}t}.
\label{g2_short}
\end{equation}
(ii) On ``long'' time scale ($t \gtrsim 20$~ns),  the antibunching associated with luminescence of the single emitterdoes not affect the $g^{(2)}$ function~\cite{Treussart} , so that:
 \begin{equation}
 g^{(2)}(t)\simeq1+ae^{-\lambda_{2}t}.
 \label{g2_long}
\end{equation}

\begin{figure}[htbp]
\begin{center}
\includegraphics[width=\textwidth]{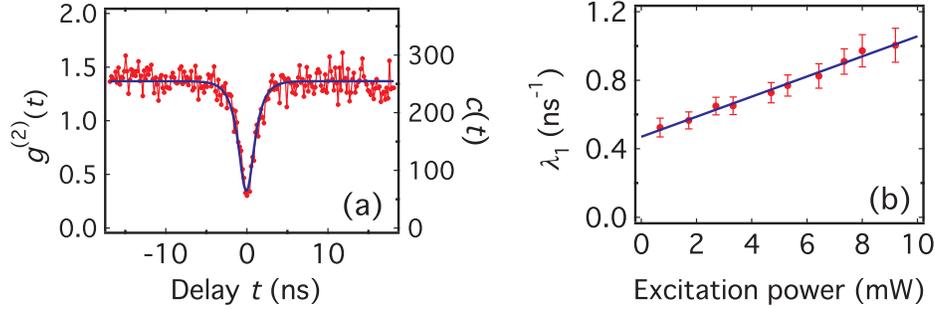}
\caption{(a) Number of photon coincidences $c(t)$ (right scale) recorded with the HBT setup for a single emitter, at short time scale ($|t|\lesssim$ 20~ns). Excitation was carried out at the maximum available cw power of 9~mW. Integration duration was $T=590$~s, $R_1\simeq37000$~counts/s, $R_2\simeq48700$~counts/s and timebin $w=0.17$~ns. Left scale: corresponding normalized intensity correlation function $g^{(2)}(t)$. The experimental data are shown as red dots, while the solid blue line represents a fit by a convolution of Eq.(\ref{g2_short}) with the measured instrumental response function. (b) Evolution of parameter $\lambda_1$ (red dots) as a function of the excitation power with linear fitting (in blue), according to Eq.(\ref{lambda1}).}
\label{fig_g2_short}
\end{center}
\end{figure} 

The normalization of the $g^{(2)}$ correlation function is inferred from the number of coincidences $c(t)$ recorded during an integration period $T$
\begin{equation}
g^{(2)}(t)=\frac{c(t)}{R_1\,R_2\, T\, w},
\end{equation}
where the factor $R_1R_2T$ corresponds to the coincidence rate for an equivalent Poissonian photon number distribution, associated with $R_{1,2}$ counts rates on each detector of the HBT setup and a time bin of width $w$. Note that this relation between $c(t)$ and $g^{(2)}(t)$ neglects any contribution from random background light emission, as we can assume from the high value of the signal-to-background ratio for the observed emitter shown in Fig.\ref{scan}(a).

The correlation function $g^{(2)}(t)$, and the associated measured number of coincidences are represented on Fig.\ref{fig_g2_short}(a) at short time scale. It shows a distinct minimum at zero delay, with a residual non-zero value caused by convolution with the instrumental response function (IRF) of the detection setup.

This function is independently measured to be a 1.2~ns FWHM Gaussian function, by correlation of Ti:Sa laser 150-fs laser pulses. Taking into account the IRF time, the measured intensity correlation function at short time-scale can be well fitted by the convolution of the IRF with the $g^{(2)}$ function given by Eq.(\ref{g2_short}) for a single emitter.

By fitting the $g^{(2)}$ function curve at  short time scale for different excitation powers, we obtain the value of $\lambda_{1}$ dependent on the excitation power (Fig.\ref{fig_g2_short}(b)). At zero excitation power, $\lambda_{1}$ is equal to the transition rate $r_{21}$, which is by definition the invert of the excited-level lifetime. For the single emitter studied in Fig.\ref{scan}(a), we get $r_{21}=0.47$ ns$^{-1}$ by extrapolating the linear fit to zero excitation power. Excited-level lifetime is then calculated to be as short as 2~ns. 

Absorption rate from the ground level is deduced from the excitation intensity dependance of  $\lambda_{1}$, by $r_{12}=\sigma I /h\nu$, where $\sigma$ is the absorption cross-section, $I$ the excitation intensity , and $\nu$ the excitation frequency. 
From $\lambda_1$ at different excitation intensities, the power-independent absorbing cross-section is evaluated to be $\sigma\simeq 1.7\times10^{-16}$~cm$^2$. 

\begin{figure} [htbp]
\begin{center}
\includegraphics[width=\textwidth]{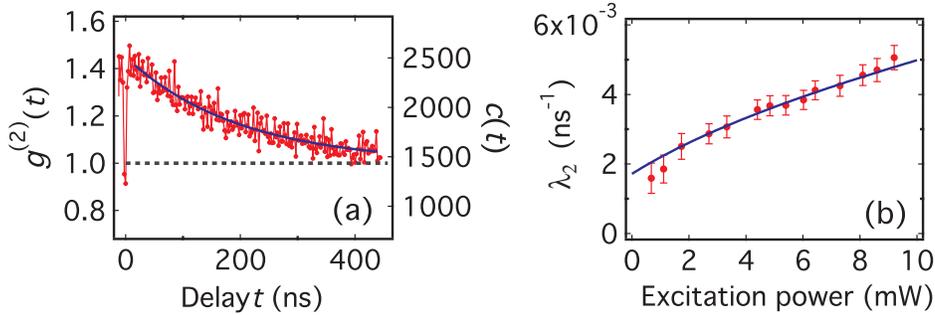}
\caption{ (a) Number of photon coincidences $c(t)$ (right scale) recorded with the HBT setup for a single emitter, on long time-scale ($|t|\gtrsim$ 20~ns). The excitation power was 9~mW. Integration duration was $T=605$~s, $R_1\simeq38600$~counts/s, $R_2\simeq33400$~counts/s and timebin $w=2.3$~ns. Left scale: corresponding normalized intensity correlation function $g^{(2)}(t)$. The experimental data are shown as red dots, while the solid blue line represents a fit by a convolution of Eq.(\ref{g2_long}) with the measured instrumental response function. The dashed line indicates the normalization level corresponding to Poissonian photon-number statistics. Note that the minimum value of $g^{(2)}$ at $t=0$ appears higher than in Fig.\ref{fig_g2_short}(a) due to choice of a larger timebin for histogram plot. (b) Evolution of parameter $\lambda_2$ (red dots) as a function of the excitation power, with fit (in blue) according to Eq.(\ref{lambda2}).}
\label{fig_g2_long}
\end{center}
\end{figure}

On long time scale, we observe $g^{(2)}(t)>1$, corresponding to a bunching effect that is due to leakage towards the ``dark'' metastable level. By fitting the $g^{(2)}(t)$ function at long time-scale neglecting the antibunching part, we obtain $\lambda_2$ as a function of the excitation intensity (Fig.\ref{fig_g2_long}(b)). Intersystem crossing rates $r_{23}$ and $r_{31}$ are evaluated by fitting $\lambda_2$ using Eq.(\ref{lambda2}), together with $r_{12}$ and $r_{21}$ deduced from the short time-scale measurements. 

Considering $r_{23}$ as a constant, we observe that the bunching effect depends on the laser power, indicating that 687~nm excitation light contributes to deshelving the metastable level. Indeed, data are well fitted by taking into account  a $r_{31}$ dependence on the excitation intensity (or equivalently on laser excitation power $P_{\rm ex}$) of the form $r_{31}=r_{31}^0(1+\beta P_{\rm ex})$, giving $r_{23}=2.75\;\mu$s$^{-1}$, $r_{31}^0=1.71\;\mu$s$^{-1}$ and $\beta=0.102\;$mW$^{-1}$. Note that such a dependance of deshelving rate with excitation intensity is also observed for single molecules~\cite{Treussart}.

With all parameters obtained above, the detected fluorescence rate is given by 
\begin{equation}
 R=\eta_{\rm det}\eta_{\rm Q}\,\frac{r_{21}}{(r_{21}/r_{12}+r_{23}/r_{31}+1)},
 \label{sat}
\end{equation}
where $R$ is the sum of count rates on the two APDs in the HBT setup. Parameters $\eta_{\rm det}$ and $\eta_{Q}$ represent the overall detection efficiency and the photoluminescence quantum yield, respectively.  With all the decay rates obtained following the fitting procedure of $\lambda_1$ and $\lambda_2$ intensity dependence, we obtain $\eta_{\rm det}\eta_{\rm Q}=4.2\times 10^{-4}$.

\begin{figure}[htbp]
\begin{center}
\includegraphics[width=0.6\textwidth]{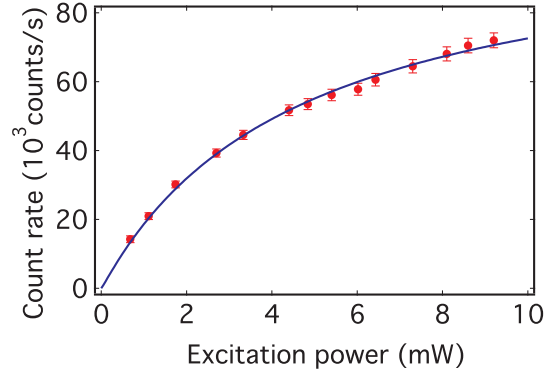}
\caption{Photoluminescence intensity measurement (plain circles with error bars) corresponding to the sum of the counting rates on the two detectors of the HBT setup, vs. excitation laser power. The line is a fit to the data by Eq.(\ref{sat}), taking into account the values of rates $r_{mn}$ determined from measurements of parameters $\lambda_1$ and $\lambda_2$. The fit is therefore achieved with a single free parameter left, {\it i.e.} the overall efficiency $\eta_{\rm det}\times\eta_{\rm Q}$.}
\label{fig_sat}
\end{center}
\end{figure}

In our experimental setup, $\eta_{\rm det}$ is limited by the collection efficiency of the microscope objective (microscope objective with NA$=0.95$ working in air without immersion oil, $\eta_{\rm col}\simeq2.6\%$), sphericity aberration ($\eta_{\rm ab}\simeq20\%$), transmittance of the objective ($\eta_{\rm trans}\simeq80\%$), transmittance of the optics (lens, dichroic mirror and filters, $\eta_{\rm opt}\simeq35\%$), and quantum efficiency of the silicon APDs in the near-infrared region ($\eta_{\rm APD}\simeq55\%$). Then the total detection efficiency is $\eta_{\rm det}= \eta_{\rm col }\times\eta_{\rm ab} \times \eta_{\rm trans} \times \eta_{\rm opt} \times\eta_{\rm APD} \simeq 8\times 10^{-4}$. We finally infer a rough estimate of the photoluminescence quantum yield as $\eta_{Q}\simeq (52\pm20)\%$. The quantum yield obviously needs a more precise evaluation. Our estimation however indicates that the measured short excited-level lifetime is presumably associated to strong radiative oscillator strength, and not to efficient photoluminescence quenching through fast non-radiative decay processes.

\section{Conclusion}
We have studied, at the single-emitter level, the photophysical properties of a near-infrared emitting colour centre, probably of the Nickel-Nitrogen related type. Assuming a three energy levels model, we have measured the key parameters associated to its photoluminescence using the intensity   
correlation function both at ``short'' and ``long'' time-scale, these two limits being compared to the emitter excited-level lifetime.
The studied single diamond colour centre has very favorable properties for applications in single-photon quantum information processing and quantum cryptography. A complementary study will be to determine the optimal excitation wavelength, in order to design a customized pulsed excitation laser to set up an efficient triggered single-photon source.

\section*{Acknowledgements}
This work was supported by an ``AC Nanosciences" grant from Minist\`ere de la Recherche,  by Institut Universitaire de France, and by Institut d'Alembert (ENS Cachan, IFR 121). This work was also partly funded by EADS/Corporate Research Centre France (Suresnes, France) under contract n$^\circ $ 74166.
E Wu is also affiliated with ``Key State Laboratory of Optical and Magnetic Resonance Spectroscopy'', as part of the joint PhD program between \'Ecoles Normales Sup\'erieures and East China Normal University.
 \end{document}